\documentclass[12pt]{article}
\usepackage[cp1250]{inputenc}
\usepackage[verbose,a4paper,tmargin=0.5in,lmargin=1in,rmargin=1in]{geometry}
\usepackage[pdftex]{graphicx}
\usepackage{amsfonts}
\usepackage{indentfirst}
\usepackage{latexsym}
\usepackage{amsmath}
\usepackage{amssymb}
\usepackage{psfrag}
\usepackage{eufrak}
\usepackage[mathscr]{eucal} 

\renewcommand{\theequation}{\arabic{section}.\arabic{equation}}

\title{All ASD complex and real 4-dimensional Einstein spaces with $\Lambda \ne 0$ admitting a nonnull Killing vector.}

\author{$\textrm{Adam Chudecki}^{*}$}

\begin{document}

\maketitle

$*$ Center of Mathematics and Physics, Lodz University of Technology, 
\newline
$\ \ \ \ \ $ Al. Politechniki 11, 90-924 Łódź, Poland, adam.chudecki@p.lodz.pl
\newline
\newline
\newline
\textbf{Abstract}. 
\newline
Anti-self-dual (ASD) 4-dimensional complex Einstein spaces with nonzero cosmological constant $\Lambda$ equipped with a nonnull Killing vector are considered. It is shown, that any conformally nonflat metric of such spaces can be always brought to a special form and the Einstein field equations can be reduced to the Boyer-Finley-Plebański equation (Toda field equation). Some alternative form of the metric are discussed. All possible real slices (neutral, Euclidean and Lorentzian) of ASD complex Einstein spaces with $\Lambda \ne 0$ admitting a nonnull Killing vector are found.


\section{Introduction}

The present paper is the second part of the work devoted to the anti-self-dual (ASD) complex Einstein spaces with $\Lambda \ne 0$ equipped with a Killing vector. In the previous part \cite{biblio_1} the general analysis and the way of classifying such spaces have been studied. It appeared, that such spaces can be divided into two classes. One class consists of the spaces which admit a nonnull Killing vector while the second contains the spaces with a null Killing vector. The case with a null Killing vector has been solved completely. Corresponding metric appeared to be of the type $[-] \otimes [\textrm{N}]$. ASD Einstein spaces with $\Lambda \ne 0$ equipped with a nonnull Killing vector (which we abbreviate by $\mathcal{H}^{\Lambda}_{K}$-spaces) admit all possible Petrov-Penrose types. The Einstein field equations in this case can be reduced to the Boyer-Finley-Plebański equation (BFP-equation or Toda field equation). 

However, in \cite{biblio_1} at least two important tasks have not been finalized. The reduction of the heavenly equation to the BFP-equation has been presented, but the reverse reduction has been not demonstrated. What is more important, the way how to reduce the metric to the form which depends on a single function which fulfills the BFP-equation has not been shown in \cite{biblio_1}. One of the aims of this paper is to fill this gap. Main result of the section 2 (Theorem 2.3) shows in details how to pass from Plebański - Robinson - Finley coordinates and the corresponding form of the metric ($W$-formalism) to the LeBrun coordinates in which the field equations are reduced to the BFP-equation ($U$-formalism). This reduction is the original result of the present paper. Moreover, simple analysis of the proof of the Theorem 2.3 shows that the reverse pass from $U$-formalism to $W$-formalism is fairly natural. 

An intermediate step between $W$-formalism and $U$-formalism is the so-called $\Sigma$-formalism. Surprisingly, it seems that the $\Sigma$-formalism provides us with the most suitable form of the metric of the complex $\mathcal{H}^{\Lambda}_{K}$-spaces. The field equations in $\Sigma$-formalism can be reduced to the form (\ref{ostateczne_rownanie}) and this equation, although still strongly nonlinear, seems to be more plausible then the BFP-equation. 

Another aim which has not been achieved in \cite{biblio_1} is the way of obtaining the real slices of the complex $\mathcal{H}^{\Lambda}_{K}$-spaces. [However, it is worth to note that the case with the null Killing vector has the natural ultrahyperbolic (neutral) slice and the corresponding metric is the most general metric of the 4-dimensional, globally Osserman spaces with nonzero curvature scalar admitting the null Killing vector \cite{biblio_1}]. The real $\mathcal{H}^{\Lambda}_{K}$-spaces have been considered by LeBrun \cite{biblio_5}, Przanowski \cite{biblio_3}, Tod \cite{biblio_4} and H\"ogner \cite{biblio_2}. Especially interesting geometrical approach based on an almost-complex structures can be found in \cite{biblio_4, biblio_2}. This approach allows to find the general forms of the metrics and to reduce the field equations to the BFP-equation.

Of course, all possible real $\mathcal{H}^{\Lambda}_{K}$-spaces are hidden inside the complex $\mathcal{H}^{\Lambda}_{K}$-spaces. The fact that real $\mathcal{H}^{\Lambda}_{K}$-metrics are known gives us the rare opportunity to examine possible techniques of obtaining real metrics from the complex ones. This aim has been achieved quite easily. We present all possible real slices of the holomorphic metric of the complex $\mathcal{H}^{\Lambda}_{K}$-spaces. The corresponding transformations, which bring us to the real slices are, in fact, quite obvious and natural (compare (\ref{trans_1})-(\ref{trans_2}), (\ref{trans_3}) and (\ref{lorentzowskie_transformacje})). The simplicity of these transformations raises hopes for the positive results in the so-called \textsl{Plebański programme}, namely: how to generate real solutions from the complex ones. Although the results of the present work concern mainly the unphysical signatures $(++--)$ and $(++++)$, they use techniques which succeeded in obtaining Lorentzian slices \cite{biblio_6}!

We believe, that our paper is the next step towards better understanding the role of complex spacetimes in general relativity. Moreover, sensible using of the techniques of obtaining real neutral and Euclidean slices could be helpful in more advanced problem concerning looking for the Lorentzian slices.

Our paper is organized as follows.

In section 2 some properties of the null and nonnull Killing vectors in Einstein spaces with $\Lambda \ne 0$ are considered. A few theorems are formulated and these theorems are generalization of the theorems presented earlier in \cite{biblio_1}. They propose the invariant criterion which can be used to distinguish null and nonnull Killing vectors. Then the detailed properties of the nonnull Killing vectors in $\mathcal{H}^{\Lambda}_{K}$-spaces are considered. These properties have not been recognized earlier in \cite{biblio_1}. Especially interesting are relations between the nonnull Killing vector, congruences of the null strings generated by this vector and Sommers vector which describes the optical properties of these null strings. However, the main result of section 2 is the Theorem 2.3, in which it is proved that the metric of the complex $\mathcal{H}^{\Lambda}_{K}$-spaces can be brought to some special form (\ref{metryka_w_postaci_Hognera}).

Sections 3, 4 and 5 are devoted to searching for the real slices of the metric obtained in section 2. We consider the neutral $(++--)$, Euclidean $(++++)$ and Lorentzian $(+++-)$ slices of the complex metric (\ref{metryka_w_postaci_Hognera}). All possible real slices are found and it is shown that in all the cases (except the Lorentzian case which can be solved completely) we arrive at the different versions of the BFP-equation.

Concluding remarks end the paper. In Appendix A we give some remarks about complex de Sitter spacetimes in various formalisms.


\setlength\arraycolsep{2pt}
\setcounter{equation}{0}

\section{Complex case}

\subsection{Killing equations and their integrability conditions in Einstein spaces}

Thorough analysis of the Killing equations and their integrability conditions for the (isometric, homothetic and conformal) Killing vectors and spinors has been presented in 
\cite{new_1}. We only remind here, that Killing equations in spinorial form for the isometric Killing vector $K_{C}^{\ \; \dot{D}}$ read
\begin{equation}
\label{ogolne_rownanie_Killinga_spinnorowo}
\nabla_{A}^{\ \; \dot{B}} K_{C}^{\ \; \dot{D}} = l_{AC} \in^{\dot{B}\dot{D}} + l^{\dot{B}\dot{D}} \in_{AC} 
\end{equation}
where $l_{AB}$ and $l_{\dot{A}\dot{B}}$ are symmetric spinors
\begin{equation}
\label{definition_ofthe_spinorrs_l}
l_{AB} := \frac{1}{2} \nabla_{(A}^{\ \ \dot{N}} K_{C)\dot{N}}
 \ , \ \ \ 
 l^{\dot{A}\dot{B}} = \frac{1}{2} \nabla^{N ( \dot{A}} K_{N}^{ \ \; \dot{B})}
\end{equation}
The integrability conditions of (\ref{ogolne_rownanie_Killinga_spinnorowo}) in Einstein spaces with $\Lambda$ ($C_{AB\dot{C}\dot{D}}=0$, $R=-4 \Lambda$) has the form
\begin{subequations}
\begin{eqnarray}
\label{integrability_c_L}
&& \nabla_{R}^{\ \; \dot{A}} l_{ST} + \frac{2}{3} \Lambda \in_{R(S}K_{T)}^{\ \ \dot{A}}  + 2 \, C_{RST}^{\ \ \ \ \; N} K^{\ \; \dot{A}}_{N} =0 \ \ \ \ \ \ \ \ 
\\
\label{integrability_c_L_dot}
&& \nabla^{A}_{\ \dot{R}} l_{\dot{S}\dot{T}} + \frac{2}{3} \Lambda  \in_{\dot{R} ( \dot{S}}K^{A}_{\ \ \dot{T})} + 2 \, C_{\dot{R}\dot{S}\dot{T}}^{\ \ \ \ \; \dot{N}} K^{A}_{\ \; \dot{N}}=0 \ \ \ \ 
\end{eqnarray}
\end{subequations}
Although we are going to analyze the ASD-metrics admitting nonnull Killing vector, we formulate theorems, which are valid without anti-self-duality assumption. These theorems are generalization of the theorems presented in \cite{biblio_1}.
\newline
\newline
\textsl{\textbf{Lemma 2.1}}
\newline
If $K_{A\dot{B}}$ is a Killing vector in 4-dimensional complex Einstein spaces with $\Lambda \ne 0$, then the spinors $l_{AB}$ and $l_{\dot{A}\dot{B}}$ defined by (\ref{definition_ofthe_spinorrs_l}) are nonzero.
\newline
\textsl{\textbf{Proof}}
\newline
Assume that $l_{AB}=0$. Then contracting (\ref{integrability_c_L}) with $\in^{RS}$ one gets $\Lambda K_{A\dot{B}} = 0$ what is contradiction. Analogously we prove, that $l_{\dot{A}\dot{B}} \ne 0$. $\blacksquare$
\newline
\newline
The next lemma follows from the general results on null Killing vectors presented in \cite{biblio_6}, although it has not been formulated there explicitly. 
\newline
\newline
\textsl{\textbf{Lemma 2.2}}
\newline
Let $K_{A\dot{B}}$ be a null Killing vector in 4-dimensional Einstein space. Then both undotted and dotted invariants $l_{AB}l^{AB}$ and $l_{\dot{A}\dot{B}} l^{\dot{A}\dot{B}}$ vanish: $l_{AB}l^{AB} = l_{\dot{A}\dot{B}} l^{\dot{A}\dot{B}} = 0$. 
\newline
\textsl{\textbf{Proof}}
\newline
Because Killing vector is null, it can be presented in the form $K_{A\dot{B}} = \mu_{A}\nu_{\dot{B}}$. Putting this into the Killing equations (\ref{ogolne_rownanie_Killinga_spinnorowo}) we find that spinors $\mu_{A}$ and $\nu_{\dot{A}}$ satisfy the equations
\begin{equation}
\label{wstegi_w_wektorze_zerowym}
\nabla_{A \dot{B}} \mu_{C} = X_{A \dot{B}} \, \mu_{C} + \in_{AC} M_{\dot{B}} \ , \ \ \ 
\nabla_{A \dot{B}} \nu_{\dot{C}} = -X_{A \dot{B}} \, \nu_{\dot{C}} + \in_{\dot{B}\dot{C}} N_{A}
\end{equation}
Moreover
\begin{equation}
\label{spinory_dla_Killinga_zerowego}
l_{AB} = \mu_{(A } N_{B)} \ , \ \ \ l_{\dot{A}\dot{B}} = \nu_{(\dot{A}} M_{\dot{B})} 
\end{equation}
and Killing equations reduce to the single condition
\begin{equation}
\label{Row_Kill_zred}
\mu_{A}N^{A} + \nu_{\dot{A}}M^{\dot{A}}=0
\end{equation}
According to complex version of Goldberg-Sachs theorem equations (\ref{wstegi_w_wektorze_zerowym}) are equivalent to the statement, that the space is algebraically special on both sides and spinors $\mu_{A}$ and $\nu_{\dot{A}}$ are, respectively,  undotted and dotted multiple Penrose spinors
\begin{equation}
\label{algebraiczna_degg}
C_{ABCD} \mu^{A}\mu^{B}\mu^{C}=0 = C_{\dot{A}\dot{B}\dot{C}\dot{D}} \nu^{\dot{A}} \nu^{\dot{B}} \nu^{\dot{C}}
\end{equation}
Putting $K_{A\dot{B}} = \mu_{A}\nu_{\dot{B}}$, (\ref{spinory_dla_Killinga_zerowego}) and (\ref{wstegi_w_wektorze_zerowym}) into (\ref{integrability_c_L}), then contracting this with $\mu^{S}\mu^{T}$ and using (\ref{algebraiczna_degg}) we obtain $\mu^{T}N_{T} M^{\dot{A}}=0$. Combining it with (\ref{Row_Kill_zred}) we find that $N_{T}=N \mu_{T}$ and $M_{\dot{T}} = M \nu_{\dot{T}}$. Finally
\begin{equation}
l_{AB} = N \mu_{A}\mu_{B} \ , \ \ \ l_{\dot{A}\dot{B}} = M \nu_{\dot{A}} \nu_{\dot{B}}
\end{equation}
Obviously $l_{AB}l^{AB} = l_{\dot{A}\dot{B}} l^{\dot{A}\dot{B}} = 0$. $\blacksquare$
\newline
\newline
The inverse lemma holds true, but it involves a little stronger assumptions. Indeed, we have
\newline
\newline
\textsl{\textbf{Lemma 2.3}}
\newline
In Einstein spaces with $\Lambda \ne 0$ isometric Killing vector is null, if $l_{AB}l^{AB}=0$ or $l_{\dot{A}\dot{B}} l^{\dot{A}\dot{B}}=0$.
\newline
\textsl{\textbf{Proof}}
\newline
Let $l_{AB}l^{AB}=0$ so $l_{AB} = \mu_{A} \mu_{B}$. Putting this into (\ref{integrability_c_L}), contracting with $K_{Z \dot{A}}$ and using the relation
\begin{equation}
K_{N}^{\ \ \dot{A}} K_{Z  \dot{A}} = \underbrace{K_{(N}^{\ \ \dot{A}} K_{Z)  \dot{A}}}_{\equiv 0} + \frac{1}{2} \epsilon_{NZ} K^{A \dot{A}}K_{A \dot{A}} = \frac{1}{2} \epsilon_{NZ} K \ , \ \ \ K:= K^{A \dot{A}}K_{A \dot{A}}
\end{equation}
we obtain
\begin{equation}
\label{rownanie_war_cal_zmienione}
K_{Z \dot{A}} \nabla_{R}^{\ \ \dot{A}} (\mu_{S}\mu_{T}) - K C_{RSTZ} + \frac{\Lambda}{3} \in_{R(S} \in_{T)Z} K = 0
\end{equation}
Contracting (\ref{rownanie_war_cal_zmienione}) with $\mu^{S} \mu^{T}$ we arrive to the condition
\begin{equation}
K (3 \mu^{S} \mu^{T} C_{RSTZ} + \Lambda \mu_{R} \mu_{Z} ) = 0
\end{equation}
so if $C_{RSTZ} \mu^{R}\mu^{S}\mu^{T} \ne 0$ (self-dual curvature is algebraically general) Killing vector is null. However if $3 \mu^{S} \mu^{T} C_{RSTZ} + \Lambda \mu_{R} \mu_{Z}  = 0$ then the self-dual curvature is algebraically special and spinor $\mu_{A}$ is undotted multiple Penrose spinor. Then from the complex Goldberg-Sachs theorem we know that the spinor $\mu_{A}$ satisfies the null string equation $\mu^{R}\mu^{S} \nabla_{R \dot{A}} \mu_{S}=0$. In this case contraction of (\ref{rownanie_war_cal_zmienione}) with $\mu^{S} \mu^{R}$ gives $K\Lambda = 0$. If cosmological constant is nonzero, then the Killing vector must be again null $K=0$. $\blacksquare$
\newline
\newline
We close this subsection by formulating a theorem which follows directly from the lemmas 2.2 and 2.3.
\newline
\newline
\textsl{\textbf{Theorem 2.1}}
\newline
Let $K_{A\dot{B}}$ be an isometric Killing vector in 4-dimensional Einstein space with $\Lambda \ne 0$ and the spinors $l_{AB}$ and $l_{\dot{A}\dot{B}}$ are defined by (\ref{definition_ofthe_spinorrs_l}). Then the following statements are equivalent
\newline
$(i)$ $K_{A \dot{B}}$ is a null vector
\newline
$(ii)$ $l_{AB}l^{AB}=0$ 
\newline
$(iii)$ $l_{\dot{A}\dot{B}}l^{\dot{A}\dot{B}}=0$ 
\newline
$\blacksquare$
\newline
\newline
From Theorem 2.1 it follows, that in Einstein spaces with $\Lambda \ne 0$ nonnull Killing vectors are characterized by the invariants $l_{AB}l^{AB}$ and $l_{\dot{A}\dot{B}} l^{\dot{A}\dot{B}}$ both being necessarily nonzero:
\begin{equation}
\nonumber
K_{A\dot{B}}K^{A\dot{B}} \ne 0 \Longleftrightarrow (l_{AB}l^{AB} \ne 0 \ \ \textrm{  and  } \ \  l_{\dot{A}\dot{B}}l^{\dot{A}\dot{B}} \ne 0 )
\end{equation}

\subsection{Nonnull Killing vectors in ASD Einstein spaces}

Now we analyze the nonnull Killing vectors in anti-self-dual Einstein spaces with cosmological constant ($C_{ABCD}=0$, $\Lambda \ne 0$). As it follows from Theorem 2.1, $l_{AB}l^{AB} \ne 0$ and $l_{\dot{A}\dot{B}}l^{\dot{A}\dot{B}} \ne 0$. We decompose the spinor $l_{AB}$ according to the formula
\begin{equation} 
\label{decomposition_of_lab_spinor}
l_{AB} = m_{(A}n_{B)} \ \ \ , \ m^{A}n_{A} \ne 0
\end{equation} 
Putting this into the equations (\ref{integrability_c_L}) we get the conditions
\begin{subequations}
\begin{eqnarray}
\label{struna_zerowa_SD}
&& \nabla_{A \dot{B}} m_{B} = Z_{A\dot{B}} \, m_{B} + \in_{AB} M_{\dot{B}}
\\
\label{struna_zerowa_ASD}
&& \nabla_{A \dot{B}} n_{B} = -Z_{A\dot{B}} \, n_{B} + \in_{AB} N_{\dot{B}}
\\
\label{postac_wektora_Killinga}
&& \frac{2}{3} \Lambda \, K_{A\dot{B}}  = -m_{A} N_{\dot{B}} - n_{A} M_{\dot{B}} \ , \ \ \ M^{\dot{N}} N_{\dot{N}} \ne 0
\end{eqnarray}
\end{subequations}
where $Z_{A\dot{B}}$, $M_{\dot{B}}$ and $N_{\dot{B}}$ are some spinors. Equations (\ref{struna_zerowa_SD}) and (\ref{struna_zerowa_ASD}) show that both spinors $m_{A}$ and $n_{A}$ form the null strings what means that the 2-dimensional holomorphic distributions
\begin{eqnarray}
&&\mathcal{D}_{m^{A}} := \{ m_{A} a_{\dot{B}}, m_{A} b_{\dot{B}} \} \ , \ \ \ a_{\dot{A}}b^{\dot{B}} \ne 0
\\
&&\mathcal{D}_{n^{A}} := \{ n_{A} a_{\dot{B}}, n_{A} b_{\dot{B}} \} 
\end{eqnarray}
are integrable in the Frobenius sense. [Remark: Distribution $\mathcal{D}_{m^{A}}$ is integrable, iff $m^{A}m^{B} \nabla_{A \dot{M}} m_{B} = 0$ what is satisfied by virtue of (\ref{struna_zerowa_SD}). Analogously, distribution $\mathcal{D}_{n^{A}}$ is integrable, iff $n^{A}n^{B} \nabla_{A \dot{M}} n_{B} = 0$ what is also satisfied by virtue of (\ref{struna_zerowa_ASD})]. The integral manifolds of these distributions constitute the congruences of self-dual null strings, i.e. holomorphic, totally null and totally geodesic 2-surfaces. [In Penrose nomenclature, these surfaces are called \textsl{$\alpha$-planes}]. $M_{\dot{A}}$ and $N_{\dot{A}}$ describe the expansion of these congruences. Both $M_{\dot{A}}$ and $N_{\dot{A}}$ are necessarily nonzero. Indeed, there are no nonexpanding congruences of the self-dual null strings in ASD Einstein space if the cosmological constant $\Lambda$ is nonzero. 

$Z_{A\dot{B}}$ are components of the \textsl{Sommers vector}. Structure of the Killing equations implies the interesting property of the congruences of the null strings generated by the distributions $\mathcal{D}_{m^{A}}$ and $\mathcal{D}_{n^{A}}$: as it follows from (\ref{struna_zerowa_SD})-(\ref{struna_zerowa_ASD}) they both have the same Sommers vector. [For detailed analysis of the optical properties of the congruences of null strings see \cite{new_2}].

Inserting (\ref{decomposition_of_lab_spinor}) and (\ref{struna_zerowa_SD})-(\ref{postac_wektora_Killinga}) into (\ref{ogolne_rownanie_Killinga_spinnorowo}) one gets
\begin{subequations}
\begin{eqnarray}
\label{warunek_na_espansje_M}
&&\nabla_{A \dot{B}} M_{\dot{C}} = Z_{A\dot{B}} \, M_{\dot{C}} + m_{A} \Big( T_{\dot{B}\dot{C}} - \frac{\Lambda}{3} \in_{\dot{B}\dot{C}} \Big)
\\
&&\nabla_{A \dot{B}} N_{\dot{C}} =- Z_{A\dot{B}} \, N_{\dot{C}} + n_{A} \Big( -T_{\dot{B}\dot{C}} - \frac{\Lambda}{3} \in_{\dot{B}\dot{C}} \Big)
\\
\label{zwiazek_lzT}
&&-\frac{2}{3} \Lambda \, l^{\dot{A}\dot{B}} = 2 N^{(\dot{A}}M^{\dot{B})} + m^{B}n_{B} \, T^{\dot{A}\dot{B}}
\end{eqnarray}
\end{subequations}
where $T^{\dot{A}\dot{B}} = T^{(\dot{A}\dot{B})}$ is some symmetric spinor related to the Sommers vector by (\ref{wlasnosci_T}) 

Now we point out some properties of the nonnull Killing vector and Sommers vector in ASD Einstein spaces with $\Lambda \ne 0$. Acting on (\ref{struna_zerowa_SD}) and (\ref{struna_zerowa_ASD}) with $\nabla_{F \dot{F}}$ and using Ricci identities for one-index spinors we find that
\begin{subequations}
\begin{eqnarray}
\label{wlasnosci_Sommersa}
&&\nabla_{(M}^{\ \ \; \dot{N}} Z_{A) \dot{N}}= 0 
\\
\label{wlasnosci_T}
&&T^{\dot{A}\dot{B}} = \nabla^{N (\dot{A}} Z_{N}^{\ \; \dot{B})} 
\end{eqnarray}
\end{subequations}
An interesting property of $T^{\dot{A}\dot{B}}$ can be found by contracting (\ref{integrability_c_L_dot}) with $\in^{\dot{R}\dot{S}}$ and using (\ref{zwiazek_lzT})
\begin{equation}
\label{property_of_the_Tab_Maxwell}
\nabla^{A\dot{N}} T_{\dot{N}\dot{B}} = 0
\end{equation}
so the spinor $T^{\dot{A}\dot{B}}$ satisfies the right Maxwell equations without currents.

The form of the nonnull Killing vector (\ref{postac_wektora_Killinga}) proves strong relation between such vector and the spinors which describe the null strings and properties of their congruences. Nonnull Killing vector in an ASD Einstein space with $\Lambda$ can be always decomposed into the sum of two null vectors which are tangent to the congruences of null strings defined by the distributions $\mathcal{D}_{m^{A}}$ and $\mathcal{D}_{n^{A}}$. Moreover, the dotted spinors in this decomposition are exactly the spinors which describe the expansion of the congruences defined by the distributions $\mathcal{D}_{n^{A}}$ and $\mathcal{D}_{m^{A}}$, respectively.

As it follows from \cite{new_2}, any null string can be brought to the \textsl{canonical normalization} without any loss of generality. [Remark: null string is called to be in a canonical normalization, if the 2-form which is interpreted as a surface element of this null string is closed]. If we assume, that the null string defined by the distribution $\mathcal{D}_{m^{A}}$ is in canonical normalization, then the spinor $m_{A}$ satisfies the equation
\begin{equation}
\label{null_string_in_CN}
\nabla^{A}_{\ \; \dot{B}} (m_{A}m_{B})=0
\end{equation}
which is obviously stronger then the classical null string equation $m^{A}m^{B}\nabla_{A\dot{B}} m_{B}=0$. Equation (\ref{null_string_in_CN}) used in (\ref{struna_zerowa_SD}) gives
\begin{equation}
\label{zwiazek_miedzy_Sommersem_i_ekspansja}
3M_{\dot{B}} = 2m^{A} Z_{A \dot{B}} \ \ \ \Longleftrightarrow \ \ \ 3M_{\dot{B}} Z_{B}^{\ \; \dot{B}} = m_{B} \, Z_{F\dot{F}}Z^{F\dot{F}}
\end{equation}
Putting (\ref{zwiazek_miedzy_Sommersem_i_ekspansja}) into (\ref{warunek_na_espansje_M}), then using (\ref{wlasnosci_Sommersa}) and (\ref{wlasnosci_T}), after some work one can get
\begin{eqnarray}
\label{covariant_derivative_of_Sommers}
\nabla_{A}^{\ \; \dot{B}} Z_{C}^{\ \; \dot{D}} &=& \frac{2m_{(A} n_{C)}}{n_{Z}m^{Z}} \, T^{\dot{B}\dot{D}} + \frac{1}{n_{Z}m^{Z}} n_{(A}Z_{C)}^{\ \ (\dot{B}} M^{\dot{D})} - \frac{m_{(A} n_{C)}}{(n_{Z}m^{Z})^{2}} \, n^{F}Z_{F}^{\ \; (\dot{B}} M^{\dot{D})} \ \ \ \ \ 
\\ \nonumber
&& +m_{A}m_{C} P^{\dot{B}\dot{D}} + \frac{1}{2} \in_{AC} T^{\dot{B}\dot{D}} - \frac{1}{6}\in_{AC}\in^{\dot{B}\dot{D}} (3\Lambda +  Z_{F\dot{F}}Z^{F\dot{F}}  )
\end{eqnarray}
where $P^{\dot{B}\dot{D}}$ is another arbitrary symmetric spinor. Note, that only one of the null strings generated by the distribution $\mathcal{D}_{m^{A}}$ or $\mathcal{D}_{n^{A}}$ can be brought to the canonical normalization, otherwise, the decomposition (\ref{decomposition_of_lab_spinor}) is not valid anymore.

There are infinitely many congruences of self-dual null strings in ASD Einstein spaces with $\Lambda \ne 0$. Two of these congruences (those generated by the distributions $\mathcal{D}_{m^{A}}$ or $\mathcal{D}_{n^{A}}$) are related to the nonnull Killing vector via (\ref{decomposition_of_lab_spinor}). On the other hand, both these congruences have the same Sommers vector. In this sense nonnull Killing vector in $\mathcal{H}^{\Lambda}_{K}$-spaces is related to Sommers vector. Knowing this, one formulates
\newline
\newline
\textsl{\textbf{Theorem 2.2}}
\newline
Let $K_{A\dot{B}}$ be a nonnull Killing vector in 4-dimensional complex ASD Einstein space with $\Lambda \ne 0$ and $\mathcal{D}_{m^{A}}$ and $\mathcal{D}_{n^{A}}$ be the distributions associated with $K_{A\dot{B}}$ via (\ref{decomposition_of_lab_spinor}) and (\ref{struna_zerowa_SD})-(\ref{struna_zerowa_ASD}). Then the Sommers vector of the congruences defined by distributions $\mathcal{D}_{m^{A}}$ and $\mathcal{D}_{n^{A}}$ is necessarily nonnull. 
\newline
\textsl{\textbf{Proof}}
\newline
Indeed, assume, that the Sommers vector is null and the congruence of the null strings generated by the distribution $\mathcal{D}_{m^{A}}$ is in canonical normalization. Then $Z_{A\dot{B}}=Z_{A}M_{\dot{B}}$ (compare (\ref{zwiazek_miedzy_Sommersem_i_ekspansja})). $Z_{A}$ is an arbitrary, nonzero spinor here. Putting this form of the Sommers vector into (\ref{covariant_derivative_of_Sommers}) and contracting it with $m^{A}M_{\dot{D}}$ one finds
\begin{equation}
\label{pomocnicze_eguation}
M_{\dot{D}}T^{\dot{D}\dot{B}} = \Lambda M^{\dot{B}} \ \ \ \Longrightarrow \ \ \ T_{\dot{B}\dot{D}}T^{\dot{B}\dot{D}} = -2 \Lambda^{2}
\end{equation}
Then using (\ref{pomocnicze_eguation}) in (\ref{property_of_the_Tab_Maxwell}) contracted with $M_{\dot{B}}$ we obtain $\Lambda=0$ what contradicts the assumption that the Sommers vector is null. $\blacksquare$

\subsection{The metric}

In \cite{biblio_1} the general form of the holomorphic ASD metric with $\Lambda$ admitting a nonnull Killing vector has been found as
\begin{eqnarray}
\label{metrykaprzestrzeni_HH_z_ekspanja}
ds^2 &=& \phi^{-2} \bigg\{ 2\tau^{-1} (d \eta  d w - d \phi dt) + 2 \, \Big( - \phi \, W_{\eta\eta} +\frac{\Lambda}{6\tau^{2}}  \Big) 
dt^{2}
\\ \nonumber
&& \ \ \ \ \ \ \ \ \ \ +4 \left( - \phi \, W_{\eta \phi} +  \, W_{\eta}\right) dw dt
  + 2 \left(  - \phi \, W_{\phi \phi} + 2 \, W_{\phi}         \right) dw^{2} \bigg\} \ \ \ \ \ \ 
\end{eqnarray}
where $W=W(\phi, \eta, w)$ is \textsl{the key function} which has to satisfy the reduced \textsl{heavenly equation with $\Lambda$}
\begin{equation}
\label{zredukowane_HH_dla_niezerr}
 \phi \, (W_{\eta \eta}W_{\phi \phi} - W_{\eta \phi}^{2}) + 2  \, (  W_{\eta} W_{\eta \phi} -  W_{\phi}W_{\eta\eta}   )
+ \frac{1}{\tau}   W_{w\eta}   - \frac{\Lambda}{6 \tau^{2}}  W_{\phi \phi}    = 0 \ \ \ \
\end{equation}
[Remark: $\tau$ is arbitrary nonzero complex constant and it can be chosen as convenient]. To obtain (\ref{zredukowane_HH_dla_niezerr}) we employ in \cite{biblio_1} the Plebański - Robinson - Finley coordinates which are adapted to the appropriate single congruence of self-dual null strings.
\newline
The metric (\ref{metrykaprzestrzeni_HH_z_ekspanja}) admits nonnull Killing vector of the form
\begin{equation}
\label{nonnnnulll_Kil}
K=\frac{\partial}{\partial t}
\end{equation}
characterized by the invariant
\begin{equation}
\label{invvariantt_gllowny}
l_{AB}l^{AB} = -2 \Big( \frac{\Lambda}{3 \tau \phi} \Big)^{2} \ne 0
\end{equation}
However, this form of the metric is inconvenient, generally because of the quite complicated equation (\ref{zredukowane_HH_dla_niezerr}). This is why we propose 
\newline
\newline
\textsl{\textbf{Theorem 2.3}}
\newline
The metric of any complex ASD Einstein space with $\Lambda$ admitting a nonnull Killing vector $K=\partial_{Z}$ can be locally brought to the form
\begin{equation}
\label{metryka_w_postaci_Hognera}
ds^{2} = \frac{V}{T^{2}} \Big( e^{U} (dX^{2} - dY^{2}) + dT^{2} \Big) - \frac{1}{V T^{2}} (dZ + \alpha)^{2}
\end{equation}
where $(X,Y,Z,T)$ are some local coordinates,
\begin{equation}
V := \frac{3}{2} \frac{ T U_{T} -2}{\Lambda}
\end{equation}
and the 1-form $\alpha$ fulfills the equation
\begin{equation}
\label{equation_for_alpha}
-\frac{2}{3} \Lambda \, d\alpha = (e^{U})_{T} \, dX \wedge dY  - T dX \wedge dU_{Y} + T dU_{X} \wedge dY
\end{equation}
The integrability condition of Eq. (\ref{equation_for_alpha}) ($d^{2}\alpha=0$) gives the Boyer - Finley - Plebański equation for $U=U(T,X,Y)$
\begin{equation}
\label{rownanie_Tody}
(e^{U})_{TT} + U_{XX} - U_{YY} = 0
\end{equation}
and the invariant, which characterizes the Killing vector reads
\begin{equation}
\label{niezmiennik_l}
l_{AB}l^{AB} = - \frac{2}{9} \frac{\Lambda^{2}}{T^{2}}  \ne 0
\end{equation}
\newline
\textsl{\textbf{Proof}}
\newline
We divide our proof in two parts I and II.
\newline
\textbf{I. $W_{\eta \eta} \ne 0$}
\newline
\textbf{Step 1: $(\phi,\eta,t,w) \rightarrow (\varphi,\eta,\varrho,v)$ (W-formalism)} 
\newline
First we re-scale the coordinates according to
\begin{equation}
\phi = \frac{\Lambda}{6 \tau^{2}} \, \varphi \ \ \ , \ w = \frac{\Lambda}{12 \tau^{3}} \, v \ \ \ , \ t=\frac{\Lambda}{6 \tau^{2}} \, \varrho
\end{equation}
so the Killing vector is
\begin{equation}
K=\frac{6 \tau^{2}}{\Lambda} \frac{\partial}{\partial \varrho}
\end{equation}
The equation (\ref{zredukowane_HH_dla_niezerr}) reads now
\begin{equation}
\label{equation_w_pierwszym_kroku}
 \varphi (W_{\eta \eta}W_{\varphi \varphi} - W_{\eta \varphi}^{2}) + 2  (  W_{\eta} W_{\eta \varphi} - W_{\varphi}W_{\eta\eta}   )
+ 2  W_{\eta v}   -   W_{\varphi \varphi}    = 0 \ \ \ \
\end{equation}
or, equivalently
\begin{eqnarray}
&&\varphi \, dW_{\varphi} \wedge dW_{\eta} \wedge dv + 2 W_{\eta} \, dW_{\eta} \wedge d\eta \wedge dv - 2W_{\varphi} \, d\varphi \wedge d W_{\eta} \wedge dv 
\\
\nonumber
&&+ 2\, d\varphi \wedge d\eta \wedge d W_{\eta} -  dW_{\varphi} \wedge d \eta \wedge dv =0
\end{eqnarray}
The metric (\ref{metrykaprzestrzeni_HH_z_ekspanja}) reduces to
\begin{eqnarray}
\label{metric_w_pierwszym_kroku}
ds^2 &=& \varphi^{-2} \bigg\{ \frac{6}{\Lambda} \, d\eta dv - \frac{2}{\tau} \, d \varphi d \varrho + \frac{\Lambda}{3 \tau^{2}} \, (1-\varphi W_{\eta \eta} ) \, d \varrho^{2} 
\\ \nonumber
&& \ \ \ \ \ \ \ \ \ \ +\frac{2}{\tau} \left(W_{\eta} - \varphi \, W_{\eta \varphi}  \right)  dv d\varrho
  + \frac{3}{\Lambda} \left( 2 \, W_{\varphi} - \varphi \, W_{\varphi \varphi}  \right)  dv^{2} \bigg\} \ \ \ \ \ \ 
\end{eqnarray}
\newline
\newline
\textbf{Step 2: $(\varphi,\eta,\varrho,v) \rightarrow (\varphi,z,\varrho,v)$ (From W-formalism to P-formalism)} 
\newline
Note that
\begin{equation}
\label{rozzniczki}
dW = W_{\varphi} \, d\varphi + W_{\eta} \, d\eta +W_{v} \, dv \ \ \Rightarrow \ \ d (W-\eta W_{\eta}) = W_{\varphi} \, d\varphi - \eta \, dW_{\eta} + W_{v} \, dv
\end{equation}
Then we use the Legendre transformation $(\varphi,\eta,v) \rightarrow (\varphi,z,v)$, $z:=W_{\eta}$
\begin{equation}
\label{transformata_Legendrea}
2P(\varphi, z, v) + \frac{1}{2} \varphi z^{2}:= W \big(\varphi, \eta(\varphi,z,v) ,v \big) - z \eta(\varphi,z,v)
\end{equation}
This transformation has sense only, if $W_{\eta \eta} \ne 0$, the case $W_{\eta \eta}=0$ must be considered individually. After some simple calculations one finds
\begin{eqnarray}
&&2P_{\varphi} + \frac{1}{2} z^{2}= W_{\varphi}  \ \ \ , \ 2P_{z} + \varphi z= -\eta  \ \ \ , \ 2P_{v} = W_{v} \ \ \ , \ W_{\eta} = z
\\ \nonumber
&&W_{\eta \eta} = - \frac{1}{2P_{zz}+\varphi} \ \ \ , \ W_{\eta \varphi} = - \frac{2P_{\varphi z} + z}{2P_{zz}+ \varphi} \ \ \ , \ W_{\varphi \varphi} = 2 P_{\varphi \varphi} - \frac{(2P_{\varphi z} + z)^{2}}{2P_{zz}+ \varphi}
\end{eqnarray}
The equation (\ref{equation_w_pierwszym_kroku}) reduces to
\begin{equation}
\label{P_equation}
P_{\varphi\varphi}P_{zz} - P_{z\varphi}^{2} + \varphi \, P_{\varphi\varphi} - P_{\varphi} + P_{zv}=0
\end{equation}
and the metric (\ref{metric_w_pierwszym_kroku}) reads
\begin{eqnarray}
\label{metryka_w_formalizmie_P}
ds^2 &=& \varphi^{-2} \bigg\{ -\frac{6}{\Lambda} (2P_{zz}+\varphi) dz dv -\frac{6}{\Lambda} (2P_{z\varphi}+z) d\varphi dv - \frac{2}{\tau} d \varphi d \varrho 
\\ \nonumber
&& \ \ \ \ \ \ \ \ \ \ + \frac{\Lambda}{3 \tau^{2}} \Big( 1 + \frac{\varphi}{2 P_{zz} + \varphi} \Big) d \varrho^{2} + \frac{2}{\tau} \Big( z + \varphi \frac{2 P_{z \varphi} +z}{2P_{zz}+ \varphi} \Big) dv d\varrho
\\ \nonumber
&& \ \ \ \ \ \ \ \ \ \ + \frac{3}{\Lambda} \bigg( 4 P_{\varphi} -4 P_{zv} +z^{2} - 2\varphi P_{\varphi \varphi} + \varphi \frac{(2 P_{z \varphi} +z)^{2}}{2P_{zz}+ \varphi} \bigg) dv^{2}
 \bigg\} \ \ \ \ \ \ 
\end{eqnarray}
Note, that the condition $W_{\eta\eta} \ne 0$ implies $2P_{zz}+ \varphi \ne 0$.
\newline
\newline
\textbf{Step 3: $(\varphi,z,\varrho,v) \rightarrow (\varphi,\xi,\varrho,v)$ (From P-formalism to $\Sigma$-formalism)} 
\newline
The equation (\ref{P_equation}) can be written in the form
\begin{equation}
\label{equation_P_form}
dR \wedge dS \wedge dv + \varphi \, dR \wedge dz \wedge dv - R \, d\varphi \wedge dz \wedge dv + d\varphi \wedge dz \wedge dS = 0
\end{equation}
where
\begin{equation}
\label{definicje_s_r}
R:=P_{\varphi} \ \ \ , \ S:=P_{z} \ \ \ \Longleftrightarrow \ \ \ dR \wedge d \varphi \wedge dv + dS \wedge dz \wedge dv = 0
\end{equation}
Eq. (\ref{equation_P_form}) is equivalent to the equation
\begin{equation}
\omega \wedge d \omega =0 \ \ \ , \ \omega := dS - R \, dv + \varphi \, dz
\end{equation}
From the Frobenius theorem it follows that there exist the functions $U=U(\varphi,z,v)$ and $\xi=\xi(\varphi,z,v)$ such that $\omega = e^{U}  d\xi$ so
\begin{equation}
\label{rozwiazanie_na_omega}
 dS - R \, dv + \varphi \, dz= e^{U}  d\xi
\end{equation}
It is easy to note that $(\varphi,\xi,v)$ can be regarded as independent variables only if $P_{zz}+\varphi \ne 0$. But this last condition is satisfied since if one had assumed that $P_{zz}+ \varphi = 0$ then Eq. (\ref{P_equation}) would not be fulfilled. Treating $z$ as a function of $(\varphi,\xi,v)$ , the equation (\ref{rozwiazanie_na_omega}) is equivalent to the set of equations
\begin{subequations}
\begin{eqnarray}
\label{rownanie_skllad_1}
&&\partial_{\xi} (S + \varphi z) = e^{U}
\\
&&\partial_{v} (S +\varphi z)=R
\\
\label{rownanie_skllad_3}
&&\partial_{\varphi} (S + \varphi z) = z
\end{eqnarray}
\end{subequations}
from (\ref{rownanie_skllad_1}) to (\ref{rownanie_skllad_3}) it follows that
\begin{equation}
\label{zwiazek_miedzy_sigma_i_SR}
\Sigma := S+\varphi z \ \ \ \Longrightarrow \ \ \ R=\Sigma_{v} \ \ \ , \ z=\Sigma_{\varphi} \ \ \ , \ e^{U} = \Sigma_{\xi} 
\end{equation}
where $\Sigma=\Sigma (\varphi,\xi,v)$. As a consequence of the condition $P_{zz} + \varphi \ne 0$ one gets $z_{\xi} \ne 0$, so $\Sigma_{\varphi \xi} \ne 0$. Compatibility condition (\ref{definicje_s_r}) takes the form
\begin{equation}
\label{rownanie_skllad_4}
R_{\xi} + S_{\xi}z_{\varphi} - S_{\varphi}z_{\xi}=0 
\end{equation}
Putting $R$ and $S$ from (\ref{zwiazek_miedzy_sigma_i_SR}) into the (\ref{rownanie_skllad_4}) one arrives at the crucial equation
\begin{equation}
\label{ostateczne_rownanie}
\Sigma_{\xi v} + \Sigma_{\xi} \Sigma_{\varphi \varphi} = 0
\end{equation}
All derivatives of the $P$ can be expressed by the derivatives of $\Sigma$ according to
\begin{eqnarray}
\label{pochodne_P_wyrazone_przez_Sigma}
&& P_{\varphi} =\Sigma_{v} \ \ \ , \ P_{z} = \Sigma - \varphi \Sigma_{\varphi} \ \ \ , \ P_{z\varphi} = \frac{\Sigma_{\xi v}}{\Sigma_{\xi \varphi}} \ \ \ , \ P_{zz} = \frac{\Sigma_{\xi}}{\Sigma_{\xi \varphi}} - \varphi
\\
&& P_{zv} = \Sigma_{v} - \frac{\Sigma_{\xi}\Sigma_{\varphi v}}{\Sigma_{\xi \varphi}} \ \ \ , \ 
P_{\varphi \varphi} = \Sigma_{v\varphi} - \frac{\Sigma_{\xi v}\Sigma_{\varphi \varphi}}{\Sigma_{\xi \varphi}}
\end{eqnarray}
Substituting (\ref{pochodne_P_wyrazone_przez_Sigma}) into (\ref{metryka_w_formalizmie_P}) one can rearranged the metric to get the form
\begin{eqnarray}
\label{zredukowana_metryka_Lmniejsze}
ds^2 &=& \varphi^{-2} \bigg\{ - \frac{2}{\tau} \, d\varphi d \varrho + \frac{2\Lambda}{3 \tau^{2}} \frac{\Sigma_{\xi}}{\Omega_{\xi}} \, d \varrho^{2} + \frac{4}{\tau} \frac{\Sigma_{\xi}\Omega_{\varphi}}{\Omega_{\xi}} \, dv d \varrho
+\frac{6}{\Lambda} \frac{\Sigma_{\xi}\Omega_{\varphi}^{2}}{\Omega_{\xi}} \, dv^{2}
\\ \nonumber
&& \ \ \ \ \ \ \ \ \ \  - \frac{6}{\Lambda} \Omega_{\varphi} \, d \varphi dv - \frac{6}{\Lambda} \Omega_{\xi} \, d \xi dv \bigg\} \ \ \ \ \  
\end{eqnarray}
where
\begin{equation}
\Omega := 2 \Sigma - \varphi \Sigma_{\varphi}
\end{equation}
The function $\Sigma$ has to satisfy the equation (\ref{ostateczne_rownanie}) with the additional condition $\Sigma_{\varphi \xi} \ne 0$. Moreover, as a consequence of $W_{\eta\eta} \ne 0$ one obtains $\Omega_{\xi} \ne 0$.
\newline
\newline
\textbf{Step 4: $(\varphi,z,\varrho,v) \rightarrow (T,X,Y,Z)$ (From $\Sigma$-formalism to $U$-formalism and BFP equation)} 
\newline
Now we transform the coordinates according to
\begin{equation}
\varrho =  -\tau Z \ \ \ , \ \ \ 
\varphi = T \ \ \ , \ \ \ 
\xi = \frac{1}{2} (Y-X) \ \ \ , \ \ \ 
v = -\frac{1}{2} (X+Y)
\end{equation}
Moreover we define the functions $V$ and $U$
\begin{equation}
V := -\frac{3}{2 \Lambda} \frac{\Omega_{\xi}}{\Sigma_{\xi}}  
 =\frac{3}{2 \Lambda} \big( \varphi ( \ln \Sigma_{\xi} )_{\varphi} -2 \big)
= \frac{3}{2 \Lambda} \big( T \, U_{T} -2 \big)
\end{equation}
where
\begin{equation}
\label{deffiniccja_U}
U := \ln \Sigma_{\xi} \ \ \ \Longleftrightarrow \ \ \ e^{U} = \Sigma_{Y} - \Sigma_{X}
\end{equation}
and the 1-form $\alpha$
\begin{equation}
\label{definicja_alpha}
\alpha := -V \, d \varphi - \frac{3}{ \Lambda} \Omega_{\varphi} \, d v = -V\, dT + \frac{3}{2\Lambda} (\Sigma_{T} - T \Sigma_{TT}) (dX + dY)
\end{equation}
Using this in (\ref{zredukowana_metryka_Lmniejsze}) one gets
\begin{equation}
\label{metryka_w_formie_przedostatniej}
ds^{2} = \frac{V}{T^{2}} \Big( e^{U} (dX^{2} - dY^{2}) + dT^{2} \Big) - \frac{1}{V T^{2}} (dZ + \alpha)^{2}
\end{equation}
Functions $U$ and $\Sigma$ are associated by the relation (\ref{deffiniccja_U}) and by the Eq. (\ref{ostateczne_rownanie}) which in this formalism has the form
\begin{equation}
\label{rownanie_laczace_sigmaiu}
\Sigma_{TT} - U_{X} - U_{Y} = 0
\end{equation}
The Killing vector reads
\begin{equation}
K=-\frac{6 \tau}{\Lambda} \frac{\partial }{\partial Z}
\end{equation}
and can be normalized to the $K=\partial_{Z}$ by special choice of the arbitrary parameter $\tau$, namely $\tau = -(\Lambda / 6)$. Then the invariant $l_{AB}l^{AB}$ given by (\ref{invvariantt_gllowny}) takes the form (\ref{niezmiennik_l}). Acting on both sides of (\ref{definicja_alpha}) with the operator $d$ one arrives at Eq. (\ref{equation_for_alpha}). Compatibility condition of the equations (\ref{deffiniccja_U}) and (\ref{rownanie_laczace_sigmaiu}) brings us to the BFP-equation (\ref{rownanie_Tody}). Consider now the case II.
\newline
\newline
\textbf{II. $W_{\eta \eta} = 0$}
\newline
If $W_{\eta \eta}=0$ then the Legendre transformation (\ref{transformata_Legendrea}) is not valid anymore. However, condition $W_{\eta \eta}=0$ leads to the function $W$ being the first-order polynomial in $\eta$ with coefficients depending on $(\phi, w)$. Reduced heavenly equation (\ref{zredukowane_HH_dla_niezerr}) can be easily solved and the solution reads
\begin{equation}
\label{funkcja_kluczowa_dlaWetaeta}
W(\eta, \phi, w) = (f_{1} \phi + f_{2}) \eta + \frac{\tau^{2}}{\Lambda} \bigg( f_{1}^{2} + \frac{1}{\tau} \frac{d f_{1}}{d w} \bigg) \phi^{3} + \frac{3 \tau^{2}}{\Lambda} \bigg( 2f_{1}f_{2} + \frac{1}{\tau} \frac{d f_{2}}{dw} \bigg) \phi^{2} + f_{3} \phi + f_{4}
\end{equation}
where $f_{1}$, $f_{2}$, $f_{3}$ and $f_{4}$ are the arbitrary functions of the variable $w$. The metric generated by the key function (\ref{funkcja_kluczowa_dlaWetaeta}) corresponds to the complex de Sitter spacetime. However, the gauge freedom still available after bringing the Killing vector into the form $\partial_{t}$ is strong enough to gauge all the functions $f_{1}$, $f_{2}$, $f_{3}$ and $f_{4}$ to zero (compare transformation formulas given in \cite{biblio_1}). Without any loss of generality one can set $W=0$. Putting $W=0$ into (\ref{metrykaprzestrzeni_HH_z_ekspanja}) and performing the coordinates transformation
\begin{equation}
\phi = T \ , \ \ \ t=\tau Z + \frac{3 \tau}{\Lambda} T \ , \  \ \ w=-(X+Y) \ , \ \ \ \eta = \frac{3\tau}{2\Lambda} (X-Y) 
\end{equation}
we obtain the metric
\begin{equation}
\label{de_Sitter_in_U_formalism}
ds^{2} = T^{-2} \bigg( - \frac{3}{\Lambda} (dX^{2}-dY^{2} + dT^{2} ) + \frac{\Lambda}{3} dZ^{2} \bigg)
\end{equation}
This is exactly the metric (\ref{metryka_w_postaci_Hognera}) with $U=0=\alpha$, $V=-(3/ \Lambda)$. After normalizing the constant $\tau$, namely $\tau=1$ the Killing vector is $K=\partial_{Z}$ and the invariant $l_{AB}l^{AB}$ has the form (\ref{niezmiennik_l}). This completes the proof. $\blacksquare$


\setlength\arraycolsep{2pt}
\setcounter{equation}{0}

\section{Neutral slices}

\subsection{H\"ogner's results}

This case of neutral signature $(++--)$ has been solved by M. H\"ogner \cite{biblio_2}. H\"ogner has proved that the metric of ASD Einstein space of the constant curvature can be always brought to some special form which depends on the sign of the invariant $l_{AB}l^{AB}$ (see \cite{biblio_2} for details). He claims, that in the special coordinate system $(t,x,y,z)$ the metric reads
\begin{equation}
\label{ultrahyperbolic_1}
ds^{2} = \frac{V}{t^{2}} \Big( e^{U} (dx^{2} \pm dy^{2}) \mp dt^{2} \Big) - \frac{1}{V t^{2}} (dz + \alpha)^{2}
\end{equation}
where 
\begin{equation}
V := \pm \frac{tU_{t} -2}{2 \Lambda}
\end{equation}
and the 1-form $\alpha$ which has to satisfy
\begin{equation}
 2 \Lambda \, d\alpha =  (e^{U})_{t} \, dx \wedge dy - t \, dx \wedge dU_{y} \mp t \, dU_{x} \wedge dy
\end{equation}
Then the integrability condition of this last equation ($d^{2} \alpha=0$) gives the BFP equation for the $U$ function 
\begin{equation}
\label{ultrahyperbolic_2}
(e^{U})_{tt} \mp U_{xx} - U_{yy} = 0
\end{equation}
The metric (\ref{ultrahyperbolic_1}) admits the nonnull Killing vector $K=\partial_{z}$. Moreover
\begin{itemize}
\item the case $l_{AB}l^{AB}>0$ corresponds to the upper signs and it describes the scalar-flat pseudo-K\"ahler space
\item the case $l_{AB}l^{AB}<0$ corresponds to the lower signs and it describes the scalar-flat para-K\"ahler space
\end{itemize}

\subsection{Real neutral slices of the metric (\ref{metryka_w_postaci_Hognera})}

In the case of the neutral (ultrahyperbolic) signature $(++--)$ the spinor $l_{AB}$ satisfies the condition
\begin{equation}
\label{reality_of_lab}
l_{AB} = \bar{l}_{AB} \ \ \Longrightarrow \ \ m_{(A}n_{B)} = \bar{m}_{(A} \bar{n}_{B)}
\end{equation}
where bar denotes the complex conjugation. Spinors $m_{A}$ and $n_{A}$ are not defined uniquely by the (\ref{decomposition_of_lab_spinor}) and they can be always re-scaled according to formula
\begin{equation}
\label{mozliwe_przeskalowania_spinorow}
m_{A} \rightarrow  \rho m_{A} \ \ \ , \ n_{A} \rightarrow \rho^{-1} n_{A} 
\end{equation}
where $\rho$ is some complex function. Straightforward analysis of the condition (\ref{reality_of_lab}) gives two possible solutions. Using (\ref{mozliwe_przeskalowania_spinorow}) they can be brought to the form
\begin{itemize}
\item $m_{A}$ and $n_{A}$ are real $\Longrightarrow l_{AB}l^{AB}=-\frac{1}{2} (m^{A}n_{A})^{2} < 0$
\item $m_{A}$ and $n_{A}$ are complex, $m_{A} = \pm \bar{n}_{A}$, $\Longrightarrow l_{AB}l^{AB}=-\frac{1}{2} (m^{A}n_{A})^{2} > 0$
\end{itemize}

In the case, when both $m_{A}$ and $n_{A}$ are real, we find from the (\ref{postac_wektora_Killinga}), (\ref{struna_zerowa_SD})-(\ref{struna_zerowa_ASD}) and (\ref{warunek_na_espansje_M}) that
$M_{\dot{A}}$, $N_{\dot{A}}$, $Z_{\dot{A}\dot{B}}$, $T_{\dot{A}\dot{B}}$ are real. Spinors $m_{A}$ and $n_{A}$ form a real null strings and it is possible to choose the spinorial base in such a manner, that $m_{A}=(0, m_{2}), m_{2} \ne 0$. This allows to introduce the real Plebański's tetrad and solve the problem from the very beginning (all the calculations are identical, like in \cite{biblio_1}). However, if $m_{A}$ and $n_{A}$ are complex, we find $M_{\dot{A}} = \pm \bar{N}_{\dot{A}}$, $Z_{A\dot{B}} + \bar{Z}_{A \dot{B}} = 0$, $T_{\dot{A}\dot{B}} + \bar{T}_{\dot{A}\dot{B}} = 0$. There are no real null strings which can be connected with the spinor $l_{AB}$ (and Killing vector) by the conditions (\ref{struna_zerowa_SD})-(\ref{struna_zerowa_ASD}). In this case to repeat H\"ogner's results from the general description given in \cite{biblio_1} is much more difficult.

The second possible way to obtain H\"ogner's results is to consider the real slices of the complex metric (\ref{metryka_w_postaci_Hognera}). Let us consider two transformations of the metric (\ref{metryka_w_postaci_Hognera})
\begin{subequations}
\begin{eqnarray}
\label{trans_1}
\textrm{transformation 1:   } && \Lambda \rightarrow -3 \Lambda
\\
\label{trans_2}
\textrm{transformation 2:   } && T \rightarrow i T \ , \ \ \ Y \rightarrow iY \ , \ \ \ \Lambda \rightarrow -3 \Lambda
\end{eqnarray}
\end{subequations}
Performing these transformations in formulas (\ref{metryka_w_postaci_Hognera}) - (\ref{niezmiennik_l}), changing the abbreviations of the coordinates into small letters and considering all coordinates, 1-forms and functions as a real smooth objects, we arrive exactly to the H\"ogner's results (\ref{ultrahyperbolic_1}) - (\ref{ultrahyperbolic_2}).

\section{Euclidean slices}

\subsection{Tod's results}

The case of the Euclidean signature has been considered by Przanowski \cite{biblio_3} and then by Tod \cite{biblio_4}. Przanowski shows in \cite{biblio_3} that the metric of the Euclidean Einstein space with nonzero cosmological constant $\Lambda$ equipped with the nonnull Killing vector can be brought to some special form and pointed out the existence of two, essentially different classes of the Killing vectors. Tod has proved, that these two classes are, in fact, identical and he has shown that the metric can be always presented in the form 
\begin{equation}
\label{Euklidesowy_1}
 ds^{2} = \frac{V}{t^{2}} \Big( e^{U} (dx^{2} + dy^{2}) + dt^{2} \Big) + \frac{1}{V t^{2}} (dz + \alpha)^{2}
\end{equation}
where
\begin{equation}
V =  \frac{t  U_{t} -2}{4 \Lambda}
\end{equation}
and the 1-form $\alpha$ satisfies the equation
\begin{equation}
 -4 \Lambda \, d\alpha =  (e^{U})_{t} \, dx \wedge dy + t \, dx \wedge dU_{y} + t \, dU_{x} \wedge dy
\end{equation}
with the integrability condition reading
\begin{equation}
\label{Euklidesowy_22}
(e^{U})_{tt} + U_{xx} + U_{yy} = 0
\end{equation}
So once again one gets the BFP equation. The Killing vector is $K=\partial_{z}$. [Remark: In \cite{biblio_4} Tod uses slightly different symbols for the coordinates and the functions. However, after changing the symbols in (\ref{Euklidesowy_1}) - (\ref{Euklidesowy_22}) according to the formulas
\begin{equation}
V \rightarrow P \ , \ \ \ \alpha \rightarrow \theta \ , \ \ \ t \rightarrow w \ , \ \ \ U \rightarrow v \ , \ \ \  z \rightarrow \tau \ \ \ \ 
\end{equation}
we arrive exactly to the Tod's results].

\subsection{Euclidean slices of the metric (\ref{metryka_w_postaci_Hognera})}

The Euclidean reality condition can be written as
\begin{equation}
l_{AB} = \bar{l}^{AB}
\end{equation}
Solving this equation and re-defining, as before, the spinors $m_{A}$ and $n_{A}$ according to (\ref{mozliwe_przeskalowania_spinorow}), we obtain only one possible solution
\begin{itemize}
\item $m_{A}$ and $n_{A}$ are complex, $m_{A} = \pm i \bar{n}^{A}$, $\Longrightarrow l_{AB}l^{AB}>0$ 
\end{itemize}
Spinors $m_{A}$ and $n_{A}$ are complex and we do not have any real null strings. The only plausible way seems to be appropriate complex transformation and then real slice of the metric (\ref{metryka_w_postaci_Hognera}). The complex transformation that we propose here reads
\begin{equation}
\label{trans_3}
T \rightarrow iT \ , \ \ \ X \rightarrow i X \ , \ \ \ \Lambda \rightarrow 6 \Lambda 
\end{equation}
Using this transformation in the formulas (\ref{metryka_w_postaci_Hognera}) - (\ref{niezmiennik_l}), changing the abbreviations of the coordinates to small letters and replacing the analytic coordinates, 1-forms and functions by real smooth ones, we obtain Tod's results (\ref{Euklidesowy_1}) - (\ref{Euklidesowy_22}). The invariant reads now
\begin{equation}
l_{AB}l^{AB} =  \frac{8 \Lambda^{2}}{ t^{2}} >0
\end{equation}

\section{Lorentzian slices}
\label{subsekcja_lorentzowskie_ciecia}

Lorentzian slices exist only if $\bar{C}_{ABCD} = C_{\dot{A}\dot{B}\dot{C}\dot{D}}$. Because $C_{ABCD}=0$ the only possible Lorentzian metric hidden inside the metric (\ref{metrykaprzestrzeni_HH_z_ekspanja}) (or (\ref{metryka_w_postaci_Hognera})) is the metric of the type $[-] \otimes [-]$ with $\Lambda \ne 0$, i.e., the metric of de Sitter spacetime. 

As it follows from Appendix A, complex de Sitter spacetime in Plebański - Robinson - Finley coordinates corresponds to the key function $W=0$. Following the second part of the proof of the Theorem 2.3 (Case $W_{\eta \eta}=0$) we find, that de Sitter spacetimes in the $U$-formalism can be always brought to the form (\ref{metryka_w_postaci_Hognera}) with $U=0=\alpha$, i.e., to the metric (\ref{de_Sitter_in_U_formalism}). Then the equations for the 1-form $\alpha$ (\ref{equation_for_alpha}) and the BFP equation (\ref{rownanie_Tody}) are identically satisfied. 

[Remark: it is worth to note, that there exist nontrivial $U$-functions which give conformally flat solutions. For example, conformally flat solutions with the $W$ function of the form (\ref{funkcja_kluczowa_dla_deSittera}) which satisfies $W_{\eta \eta} \ne 0$, $W_{\eta \eta \eta}=0$, translated into $U$-formalism give
\begin{equation}
\label{przyklad_confrom_flat_1}
U_{T} = \frac{12b \, T + 4c}{6 b \, T^{2} + 2 c \, T -1} \ , \ \ \ b=b(X+Y) \ , \ c=c(X+Y)
\end{equation}
This equation can be integrated in all subcases. If $b$ and $c$ are constants, the BFP-equation reduces to the Liouville equation \cite{biblio_7, biblio_9} which solution is well known. 

The other family of the $U$-function which give the conformally flat solution is the equivalent of the key functions (\ref{funkcja_kluczowa_dla_deSittera}) which satisfy $W_{\eta \eta \eta} \ne 0$. In this case the $U$-function has to satisfy the following equation
\begin{equation}
\label{przyklad_confrom_flat}
6a \, e^{U} = (\partial_{Y}-\partial_{X}) \bigg( \frac{1}{T^{2}(TU_{T} - 2)^{2}} \bigg) \ , \ \ \ a=a(X+Y)
\end{equation}
Using (\ref{przyklad_confrom_flat}) with $a=\textrm{const}$, the BFP equation can be once integrated, but the full explicit solution is not known].

Using in the metric (\ref{de_Sitter_in_U_formalism}) the following coordinate transformation
\begin{equation}
\tau T = \frac{\Lambda}{6} u - v \ , \ \ \ \tau Z = \frac{u}{2} + \frac{3v}{\Lambda} \ , \ \ \ X = \frac{\Lambda \xi}{3 \tau^{2}} - \frac{\zeta}{2} \ , \ \ \ Y= -\frac{\Lambda \xi}{3 \tau^{2}} - \frac{\zeta}{2}
\end{equation}
we obtain exactly de Sitter metric in the form (\ref{appen_de_sitter_metric_simply2}). Using one more coordinate transformation (\ref{appen_zwiazek_miedzy_coordynatami}) we obtain the equivalent form of the metric of de Sitter spacetime (\ref{appen_de_sitter_metric_simply3}). Considering the coordinates $(x,y,z,t)$ in (\ref{appen_de_sitter_metric_simply3}) as a real coordinates, we obtain the real de Sitter metric in Lorentzian signature.


\section{Concluding remarks}

Our paper is a concise summary of the problem of dealing with the real $\mathcal{H}^{\Lambda}_{K}$-spaces via complex ones. The transformation leading from Plebański - Robinson - Finley coordinates to coordinates employed in \cite{biblio_4, biblio_2} is presented in details. Both coordinates systems have deep geometrical meaning. Plebański - Robinson - Finley coordinates are adapted to the congruence of the null strings and seem to be transparent in considerations in complex space-times. Coordinates used by Tod and H\"ogner are associated with the almost-complex or para-almost-complex structures. It is obvious, that the transformation considered can be treated as a bridge between these two geometrical approaches. 

The next conclusion concerns real slices of the complex metrics. Transformations (\ref{trans_1}), (\ref{trans_2}) and (\ref{trans_3}) are, in fact, based on a sensible use of imaginary unit $i$. Elsewhere, the same approach allowed us to obtain Lorentzian slices of the types $[\textrm{II}]\otimes[\textrm{II}]$ and $[\textrm{D}]\otimes[\textrm{D}]$ (compare \cite{biblio_6}). The important question arises: does the same technique can be used for more advanced problem, namely for the type $[\textrm{D}]\otimes[\textrm{any}]$ real slices? In \cite{biblio_7} the general form of the metric of the one-sided type [D] gravitational instantons has been found. The similarity between this metric and the metric (\ref{metryka_w_postaci_Hognera}) is obvious. We are going to investigate the problem of all real type $[\textrm{D}]\otimes[\textrm{any}]$ spaces via complex ones in future.

It is worth-while to mention the form of the metric of complex $\mathcal{H}^{\Lambda}_{K}$-spaces in $\Sigma$-formalism (\ref{zredukowana_metryka_Lmniejsze}). It depends on the function $\Sigma=\Sigma(\varphi,v,\xi)$ which has to satisfy the equation (\ref{ostateczne_rownanie}) with the conditions $\Sigma_{\varphi \xi} \ne 0$, $2\Sigma_{\xi}-\varphi \Sigma_{\xi \varphi} \ne 0$. It seems, that this form of metric of $\mathcal{H}^{\Lambda}_{K}$-spaces is the most plausible. It does not involve any additional 1-form $\alpha$ (like  (\ref{metryka_w_postaci_Hognera})) and any additional equation (like (\ref{equation_for_alpha})). What is the most important, any solution of the equation (\ref{ostateczne_rownanie}) gives immediately the metric of the $\mathcal{H}^{\Lambda}_{K}$-spaces. For example, Plebański and Finley in \cite{biblio_8} introduced two families of the solutions of the BFP-equation, but they did not present the corresponding metrics. These families in $\Sigma$-formalism read
\begin{subequations}
\begin{eqnarray}
\label{example_of_solution_1}
&&\Sigma = -\frac{1}{2} f_{v} \, \varphi^{2} + e^{f} (a \varphi + b) 
\\
\label{example_of_solution_2}
&&\Sigma= -\frac{1}{2} f_{v} \, \varphi^{2} + a e^{f} (\varphi + g) - g_{v} \, (\varphi + g) \ln (\varphi + g)
\end{eqnarray}
\end{subequations}
where $f=f(v)$, $g=g(v)$, $b=b(\xi)$ and $a=a(\xi)$, $a_{\xi} \ne 0$ are the arbitrary functions of their variables. The metrics corresponding to (\ref{example_of_solution_1}) and (\ref{example_of_solution_2}) in $\Sigma$-formalism can be obtain directly by putting (\ref{example_of_solution_1}) and (\ref{example_of_solution_2}) into (\ref{zredukowana_metryka_Lmniejsze}). Can the equation (\ref{ostateczne_rownanie}) be useful in finding other explicit examples of the $\mathcal{H}^{\Lambda}_{K}$-spaces? What about all algebraically degenerated $\mathcal{H}^{\Lambda}_{K}$-spaces? We are going to answer these questions soon.

Considering de Sitter spacetimes within the $U$-formalism we have faced on interesting problem, namely: find all solutions of the BFP equation which give the conformally flat, Einstein spaces with nonzero cosmological constant. In the present paper we have found two families of solutions (compare (\ref{przyklad_confrom_flat_1}) and  (\ref{przyklad_confrom_flat})), but we are still far from the full solution of the problem. It seems, that the way of "translation" the conformally flat solutions from the $W$-formalism into $U$-formalism via steps listed in the proof of Theorem 2.3 is not efficient. Only a deep analysis of the structure of metric (\ref{metryka_w_postaci_Hognera}) perhaps allows to solve the issue.
\newline
\newline
\textbf{Acknowledgments.} 

I am indebted to prof. M. Przanowski for many enlightening discussions and  his interest in this work.


\appendix
\renewcommand{\theequation}{\Alph{section}.\arabic{equation}}
\setcounter{equation}{0}
\section{Appendix. Notes on complex de Sitter spacetimes}

\subsection{Complex de Sitter spacetime in null tetrad formalism}

In this appendix we write down some forms of the metric of complex de Sitter spacetimes. [Analogous considerations about real de Sitter spacetimes can be found in \cite{biblio_10}]. The metric of any complex conformally flat spacetime can be led to the form
\begin{equation}
\label{appen_conformally_flat}
ds^{2} = \Phi^{-2} (dx^{2} + dy^{2} + dz^{2} - dt^{2})
\end{equation}
where $(x,y,z,t)$ are some complex coordinates and $\Phi=\Phi(x,y,z,t)$ is some complex function. Introducing the new complex coordinates 
\begin{equation}
\label{appen_zwiazek_miedzy_coordynatami}
\xi = \frac{1}{\sqrt{2}} (x + iy) \ , \ \ \ \zeta = \frac{1}{\sqrt{2}} (x - iy) \ , \ \ \ u = \frac{1}{\sqrt{2}} (z + t) \ , \ \ \ v = \frac{1}{\sqrt{2}} (z - t) 
\end{equation}
we arrive at the form of metric
\begin{equation}
ds^{2} = \Phi^{-2} ( 2d \xi d \zeta + 2 du dv)
\end{equation}
Choosing the null tetrad 
\begin{equation}
e^{1} = \Phi^{-1} d \xi \ , \ \ \ e^{2} = \Phi^{-1} d \zeta \ , \ \ \ e^{3} = \Phi^{-1} du \ , \ \ \ e^{4} = \Phi^{-1} dv
\end{equation}
and using the first structure equations we find the connection forms to be
\begin{eqnarray}
&& \Gamma_{42} = \Phi_{v} \, e^{1} - \Phi_{\zeta} \, e^{3} \ , \ \ \ \Gamma_{41} = \Phi_{v} \, e^{2} - \Phi_{\xi} \, e^{3}
\\ \nonumber
&& \Gamma_{31} = \Phi_{u} \, e^{2} - \Phi_{\xi} \, e^{4} \ , \ \ \ \Gamma_{32} = \Phi_{u} \, e^{1} - \Phi_{\zeta} \, e^{4}
\\ \nonumber
&& \pm \Gamma_{12} + \Gamma_{34} = \pm \Phi_{\xi} \, e^{1} \mp \Phi_{\zeta} \, e^{2} + \Phi_{u} \, e^{3} - \Phi_{v} \, e^{4}
\end{eqnarray}
From the second structure equations one obtains
\begin{equation}
C_{ABCD} = C_{\dot{A}\dot{B}\dot{C}\dot{D}}=0 \ , \ \ \ 
\frac{R}{12}  = 2 \Phi_{\zeta} \Phi_{\xi} + 2 \Phi_{v}\Phi_{u} - \Phi \Phi_{\zeta \xi} - \Phi \Phi_{uv}
\end{equation}
and
\begin{equation}
\left( -\frac{1}{2} \, C_{ab} \right) = \Phi 
\left(
\begin{array}{cccc}
\Phi_{\xi \xi} & \frac{1}{2} \, (\Phi_{\xi \zeta}-\Phi_{uv}) & \Phi_{\xi u} & \Phi_{\xi v} \\
\frac{1}{2} \, (\Phi_{\xi \zeta}-\Phi_{uv}) & \Phi_{\zeta \zeta} &
\Phi_{\zeta u} & \Phi_{\zeta v}  \\
\Phi_{u \xi} & \Phi_{u \zeta} & \Phi_{uu} & \frac{1}{2} \, (\Phi_{uv} - \Phi_{\xi \zeta})  \\
\Phi_{v \xi} & \Phi_{v \zeta} & \frac{1}{2} \, (\Phi_{uv} - \Phi_{\xi \zeta}) & \Phi_{vv}
\end{array}
\right)
\end{equation}
Demanding additionally that $C_{ab}=0 \Longrightarrow R=-4 \Lambda$ we arrive at the formulas
\begin{eqnarray}
&&\Phi = \alpha_{0} (\xi \zeta + uv) + \beta_{0} \zeta + \mu_{0} \xi + \gamma_{0} u + \delta_{0} v + \epsilon_{0} 
\\
\label{appen_warunek_na_stale}
&& \frac{\Lambda}{6} = \alpha_{0} \epsilon_{0} - \beta_{0} \mu_{0} - \gamma_{0} \delta_{0}
\end{eqnarray}
Gathering, the metric of the complex de Sitter spacetimes can be always brought to the form
\begin{equation}
\label{appen_metric_de_sitter_ostatecczna}
ds^{2} = \frac{2d \xi d \zeta + 2 du dv}{\big( \alpha_{0} (\xi \zeta + uv) + \beta_{0} \zeta + \mu_{0} \xi + \gamma_{0} u + \delta_{0} v + \epsilon_{0}  \big)^{2}}
\end{equation}
where $\Lambda$ is the cosmological constant and $\alpha_{0}$, $\beta_{0}$, $\mu_{0}$, $\gamma_{0}$, $\delta_{0}$ and $\epsilon_{0}$ are complex constants satisfying (\ref{appen_warunek_na_stale}).

There are at least two ways of simplifying the form of the metric. Changing the coordinates in (\ref{appen_metric_de_sitter_ostatecczna}) according to
\begin{equation}
\xi = \frac{\Lambda}{6 \alpha_{0}} \xi' - \frac{\beta_{0}}{\alpha_{0}} \ , \ \ \ 
\zeta = \frac{\Lambda}{6 \alpha_{0}} \zeta' - \frac{\mu_{0}}{\alpha_{0}} \ , \ \ \ 
u = \frac{\Lambda}{6 \alpha_{0}} u' - \frac{\delta_{0}}{\alpha_{0}} \ , \ \ \ 
v = \frac{\Lambda}{6 \alpha_{0}} v' - \frac{\gamma_{0}}{\alpha_{0}} 
\end{equation}
and dropping primes the metric (\ref{appen_metric_de_sitter_ostatecczna}) reads
\begin{equation}
\label{appen_de_sitter_metric_simply1}
ds^{2} = \frac{2d \xi d \zeta + 2 du dv}{\big[ 1+ \frac{\Lambda}{6} (\xi \zeta + uv) \big]^{2}} \stackrel{(\ref{appen_zwiazek_miedzy_coordynatami})}{=}  \frac{dx^{2} + dy^{2} + dz^{2} - dt^{2}}{\big[ 1+ \frac{\Lambda}{12} (x^{2} + y^{2} + z^{2} - t^{2}) \big]^{2}} 
\end{equation}
If coordinates $(x,y,z,t)$ are real, the metric (\ref{appen_de_sitter_metric_simply1}) has the Lorentzian signature $(+++-$) and it is the well known form of the metric of the Lorentzian de Sitter spacetime.

However, one can always use the tetrad gauge freedom. Detailed analysis proves that one can always set
\begin{equation}
\Phi_{\xi} = \Phi_{\zeta}=0 \ , \ \ \ \Phi_{v}=-1 \ \ \ \Longrightarrow \ \ \ \alpha_{0}=\beta_{0}=\mu_{0}=0 \ , \ \ \ \delta_{0} = -1 \ , \ \ \ \gamma_{0} = \frac{\Lambda}{6}
\end{equation}
The constant $\epsilon_{0}$ can be absorbed then into $v$ coordinate and the metric takes the form
\begin{subequations}
\begin{eqnarray}
\label{appen_de_sitter_metric_simply2}
ds^{2} &=& \frac{2d \xi d \zeta + 2 du dv}{\big[ \frac{\Lambda}{6} u -v \big]^{2}} 
\\ 
\label{appen_de_sitter_metric_simply3}&\stackrel{(\ref{appen_zwiazek_miedzy_coordynatami})}{=}&
 \frac{2( dx^{2} + dy^{2} + dz^{2} - dt^{2} )}{\big[ \frac{\Lambda}{6}  (z+t) -  (z-t) \big]^{2}}
\end{eqnarray}
\end{subequations}
The form (\ref{appen_de_sitter_metric_simply2}) of the metric of the complex de Sitter spacetime is especially useful for our purposes (see section \ref{subsekcja_lorentzowskie_ciecia}).

\subsection{Complex de Siter spacetimes in the Plebański - Robinson - Finley coordinates}

One can attack the problem of complex de Sitter spacetimes using the heavenly spaces theory. Complex de Sitter spacetimes are characterized by the conditions $C_{ABCD}=C_{\dot{A}\dot{B}\dot{C}\dot{D}}=C_{AB \dot{A}\dot{B}}=0$ and $\Lambda \ne 0$. Analysis of the curvature formulas \cite{biblio_1} proves, that the most general key function $W$ which gives de Sitter spacetime is the third-order polynomial in $\eta$ and $\phi$ and it has the form
\begin{equation}
\label{funkcja_kluczowa_dla_deSittera}
W = g_{1} \, \eta^{3} + g_{2} \, \eta^{2} \phi + g_{3} \, \eta \phi^{2} + g_{4} \, \phi^{3} + g_{5} \, \eta^{2} + g_{6} \, \eta \phi + g_{7} \, \phi^{2} + g_{8} \, \eta + g_{9} \, \phi + g_{10}
\end{equation}
where $g_{1},...,g_{10}$ are functions of $(w,t)$. Of course, the key function (\ref{funkcja_kluczowa_dla_deSittera}) has to satisfy heavenly equation with $\Lambda$, what gives some constraints on functions $g_{1},...,g_{10}$. However, simple analysis proves that without any loss of generality one can put in de Sitter spacetimes $W=0$. [We do not present here the detailed considerations, only mention, that to prove this fact it is necessary to choose some special congruence of the null strings]. The heavenly equation with $\Lambda$ is then identically satisfied. Gathering, not loosing generality but only using the freedom in choosing the congruence of the null strings, one can always brought the metric of the complex de Sitter spacetime in Plebański - Robinson - Finley coordinates to the form
\begin{equation}
\label{metrykaprzestrzeni_deSittera_Plebanski}
ds^2 = \phi^{-2} \bigg( 2\tau^{-1} (d \eta  d w - d \phi dt) + \frac{\Lambda}{3\tau^{2}}  dt^{2} \bigg)
\end{equation}
This metric admits ten Killing vectors. Inserting the key function $W=0$ into the general master equation (formula (3.16) of Ref. \cite{biblio_1}) we find these vectors as
\begin{eqnarray}
&&\partial_{t} \ , \ \ \ \partial_{w} \ , \ \ \ \partial_{\eta} \ , \ \ \ t \partial_{t} + \phi \partial_{\phi} + 2 \eta \, \partial_{\eta} \ , \ \ \ w \partial_{w} - \eta \partial_{\eta} \ , \ \ \ 
\\ \nonumber
&&
\eta \partial_{t} + \Big( \phi - \frac{\Lambda t}{3 \tau} \Big) \partial_{w} \ , \ \ \ w \partial_{t} + \Big( \phi - \frac{\Lambda t}{3 \tau} \Big) \partial_{\eta}
\\ \nonumber
&&tw \partial_{t} + w^{2} \partial_{w} + \phi w \partial_{\phi} + \Big( \phi t - \frac{\Lambda t^{2}}{6 \tau} \Big) \partial_{\eta} \ , \ \ \ t \eta \partial_{t} + \Big( t \phi - \frac{\Lambda t^{2}}{6 \tau} \Big) \partial_{w} + \eta \phi \partial_{\phi} + \eta^{2} \partial_{\eta}
\\ \nonumber
&&\Big(w \eta - \frac{\Lambda t^{2}}{6 \tau} \Big) \partial_{t} + \Big(w \phi - \frac{\Lambda tw}{3 \tau} \Big) \partial_{w} + \Big(\phi^{2} - \frac{\Lambda \phi t}{3 \tau} \Big) \partial_{\phi} + \Big( \eta \phi - \frac{\Lambda \eta t}{3 \tau} \Big) \partial_{\eta}
\end{eqnarray}
Changing the coordinates in (\ref{metrykaprzestrzeni_deSittera_Plebanski}) according to
\begin{equation}
\label{lorentzowskie_transformacje}
\tau \eta = \xi \ , \ \ \ w=\zeta \ , \ \ \ t=u \ , \ \ \ \tau \phi = \frac{\Lambda}{6} u-v
\end{equation}
we arrive at the form (\ref{appen_de_sitter_metric_simply2}).


\end{document}